\documentclass[aapm,mph,graphicx,longbibliography]{revtex4-1}
\usepackage[T1]{fontenc}
\usepackage[utf8]{inputenc}
\usepackage{graphicx}
\usepackage{amsmath, amssymb}
\usepackage{booktabs}
\usepackage{natbib}
\usepackage{caption,subcaption}
\usepackage{hyperref}
\newcommand{\SI}[2]{\ensuremath{#1\,\mathrm{#2}}}

\let\bar=\overline
\newcommand{\imax}{\ensuremath{I_{\mathrm{max}}}}
\newcommand{\imin}{\ensuremath{I_{\mathrm{min}}}}

\begin{document}

\title{Investigation of the signature of lung tissue in X-ray grating-based phase-contrast imaging}
\author{Thomas Weber}
\email{thomas.weber@physik.uni-erlangen.de}
\affiliation{ECAP - Erlangen Centre for Astroparticle Physics,\\%
   University of Erlangen-Nuremberg,\\%
   Erwin-Rommel-Str. 1, 91058 Erlangen, Germany}

\author{Florian Bayer}
\affiliation{ECAP - Erlangen Centre for Astroparticle Physics,\\%
   University of Erlangen-Nuremberg,\\%
   Erwin-Rommel-Str. 1, 91058 Erlangen, Germany}

\author{Wilhelm Haas}
\altaffiliation{ECAP - Erlangen Centre for Astroparticle Physics,\\%
   University of Erlangen-Nuremberg,\\%
   Erwin-Rommel-Str. 1, 91058 Erlangen, Germany}
\affiliation{University of Erlangen-Nuremberg,\\%
   Pattern Recognition Lab,\\%
   Martensstr. 3, 91058 Erlangen, Germany}

   \author{Georg Pelzer}
\affiliation{ECAP - Erlangen Centre for Astroparticle Physics,\\%
   University of Erlangen-Nuremberg,\\%
   Erwin-Rommel-Str. 1, 91058 Erlangen, Germany}
   
   \author{Jens Rieger}
\affiliation{ECAP - Erlangen Centre for Astroparticle Physics,\\%
   University of Erlangen-Nuremberg,\\%
   Erwin-Rommel-Str. 1, 91058 Erlangen, Germany}
   
   \author{Andr\'{e} Ritter}
\affiliation{ECAP - Erlangen Centre for Astroparticle Physics,\\%
   University of Erlangen-Nuremberg,\\%
   Erwin-Rommel-Str. 1, 91058 Erlangen, Germany}
   
   \author{Lukas Wucherer}
\affiliation{ECAP - Erlangen Centre for Astroparticle Physics,\\%
   University of Erlangen-Nuremberg,\\%
   Erwin-Rommel-Str. 1, 91058 Erlangen, Germany}
   
   \author{Jan Matthias Braun}
\affiliation{Franz-Penzoldt-Zentrum, University Hospital\\%
			 University of Erlangen-Nuremberg,\\%
 		Palmsanlage 5, 91054~Erlangen~(Germany)}
 		
   \author{J\"urgen Durst}
\affiliation{ECAP - Erlangen Centre for Astroparticle Physics,\\%
   University of Erlangen-Nuremberg,\\%
   Erwin-Rommel-Str. 1, 91058 Erlangen, Germany}
   
\author{Thilo Michel}
\affiliation{ECAP - Erlangen Centre for Astroparticle Physics,\\%
   University of Erlangen-Nuremberg,\\%
   Erwin-Rommel-Str. 1, 91058 Erlangen, Germany}
   
\author{Gisela Anton}
\affiliation{ECAP - Erlangen Centre for Astroparticle Physics,\\%
   University of Erlangen-Nuremberg,\\%
   Erwin-Rommel-Str. 1, 91058 Erlangen, Germany}

\date{\today}

\begin{abstract}
\textbf{Purpose:} Grating-based X-ray phase-contrast imaging is a promising modality increasing the soft tissue
contrast in medical imaging. In this work, the signature of lung tissue in X-ray grating-based physe-contrast imaging
is investigated.\\
\textbf{Methods:} We used a Talbot-Lau interferometer for our investigations of two C57BL/6 mice. Both underwent projection imaging and computed
tomography.\\
\textbf{Results:} The results show that the three images obtained by X-ray phase-contrast imaging show complementary anatomical
structures. Especially the dark field image allows a more-exact determination of the position of the lung in the chest
cavity.\\
\textbf{Conclusion:} Due to its sensitivity to granular structures, the dark field image may be used for the
diagnosis of lung diseases in earlier stages or without a CT scan. Furthermore, X-ray phase-contrast imaging may also
have great potential in the application of animal laboratory sciences to reduce the number of required animals used in long-term translational, toxicity, and regenerative medicine
studies.


\end{abstract}

\maketitle

\section{Introduction}
Since the discovery of X-ray radiation by W.C.~Röntgen in 1895, its attenuation properties have been used in many
fields such as material science and medicine. On the other hand, as an electromagnetic wave, X-rays are not only
absorbed when traversing matter, but also shifted in phase. To
detect this phase-shift, many techniques mainly for the use at synchrotron facilities have been
developed~\cite{
snigirev95,wilkins96,chapman97}. Another
approach which was published the last few years is the usage of a
grating interferometer~\cite{david02,momose03,weitkamp05,momose06}. Furthermore, it has already been shown
that it is possible to employ this technique at low-brilliance, medical X-ray sources~\cite{pfeiffer06} to improve the
soft-tissue contrast.

One possible medical application exploiting this ability of X-rays is lung imaging. Here, attenuation-based X-ray
imaging is the most-often-used modality to obtain morphological information. It has been shown by
\citet{kitchen04,kitchen05}, that lung tissue seems to be the ideal organ for phase-contrast imaging, due
to its small-sized alveoli.

Kitchen et al. carried out their measurements at a synchrotron facility using a high flux monoenergetic
X-ray beam and a high-resolution X-ray detector, not applicable for medical routine. In this
work, in contrast, we report on our investigations of \emph{in situ} lung tissue imaging in a grating-based
X-ray phase-contrast setup using a conventional, medical X-ray tube and a commercial X-ray flat-panel detector as it
is widely used in a medical environment.
\section{Materials and Methods}
\subsection{Basics of Talbot-Lau interferometry}
\begin{figure}
	\begin{center}
		\includegraphics[width=.8\linewidth,clip,trim=175 210 140 75]{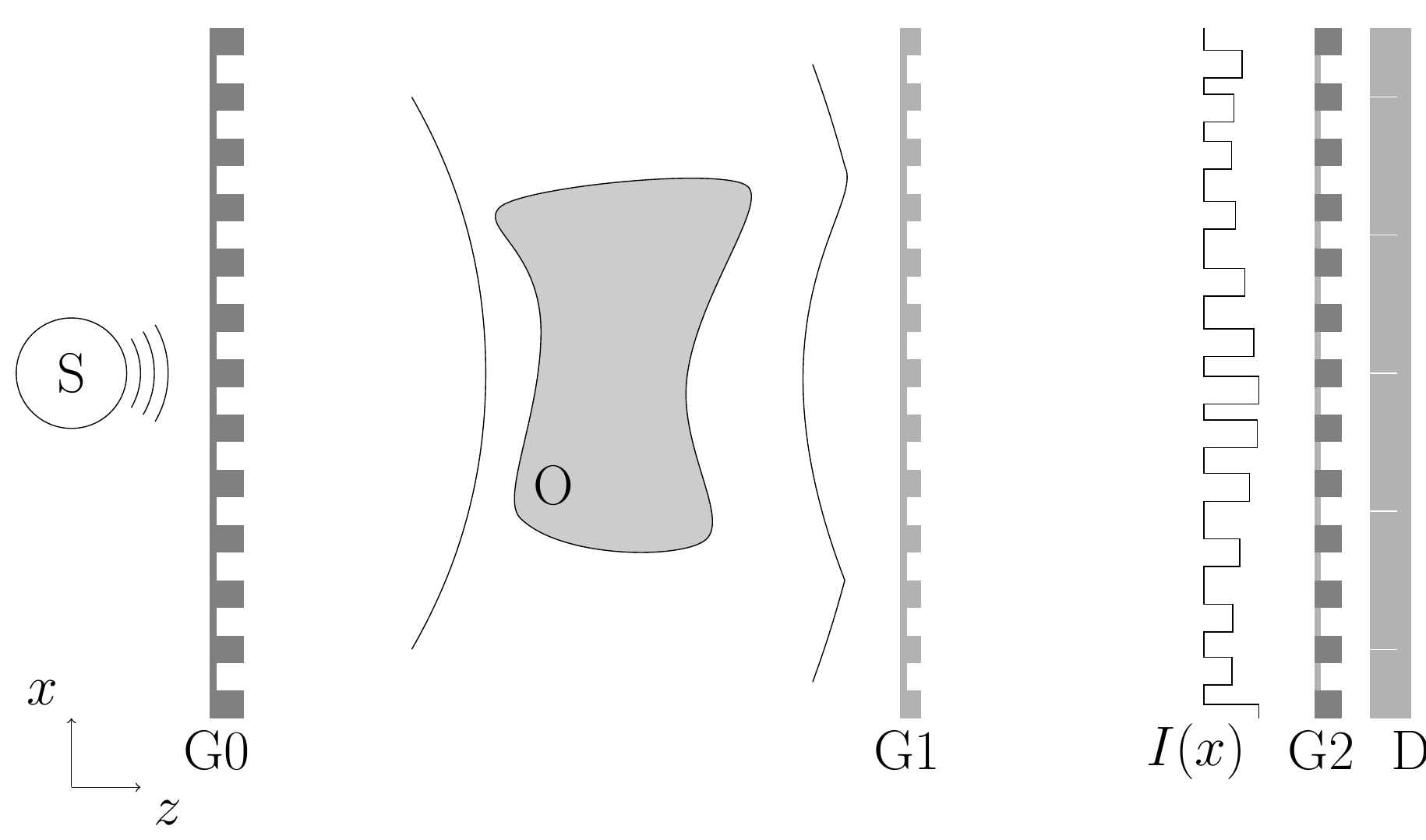}
		\caption{Sketch of the used Talbot-Lau Interferometer with X-ray source~S, source grating~G0, a sample~O,
		the phase grating~G1, the analyser grating~G2, and a pixelated detector~D. $I(x)$ illustrates the idealized,
		monoenergetic Talbot intensity pattern, produced by G1 and distorted by the object.}
		\label{fig:interferometer}
	\end{center}
\end{figure}

X-ray phase-contrast imaging can be performed with low-brilliance medical X-ray sources using Talbot-Lau
interferometry. This technique was proposed by \citet{pfeiffer06}. The basic idea of such a setup is
shown in Figure~\ref{fig:interferometer}.
A medical X-ray source produces non-coherent, polychromatic X-rays. Downstream the direction of wave
propagation a source grating (G0) ensures the spatial coherence by introducing
many virtual slit sources. The wavefronts originating from these slit sources impinge on the object,
where they become deformed depending on the object's material properties. Further towards the detection plane a phase
grating (G1) imprints a periodical phase shift on the wavefronts emitted from G0. For a
coherent, monoenergetic plane wave and a $\pi$-shifting phase grating G1 a rectangular intensity modulation occurs in
the so-called constructive fractional Talbot distances $d_T$~\cite{talbot36,suleski97} behind G1:
\begin{equation}
	d_T = \frac{\left(2n-1\right)g_1^2}{8\lambda}.\label{eq:talbot1}
\end{equation}
For a spherical wave this distances rescale due to the magnification of the setup to
\begin{equation}
	d_T^* = \frac{l d_T}{l - d_T}, \label{eq:talbot}
\end{equation}
where $g_1$ is the grating constant of the phase grating, $\lambda$ the wavelength of the X-rays, $l$ the G0 to G1
distance, and $n$ an integer number.

The resulting regular intensity pattern $I(x)$, which is in general too small to be measured directly, can be
sampled by a measurement of the intensity for a number of grating positions $p$ of the analyser grating (G2) called
``phase steps''. This technique was adapted by ~\citet{weitkamp05} from visible light
interferometry~\cite{morgan82,creath88}. When this procedure is repeated once with and once without the
sample, two intensity oscillations $I(p)$ are obtained in each pixel of the pixelated detector~(D).

\begin{figure}
	\begin{center}
	  \includegraphics[width=.8\linewidth,clip,trim=230 400 225 50]{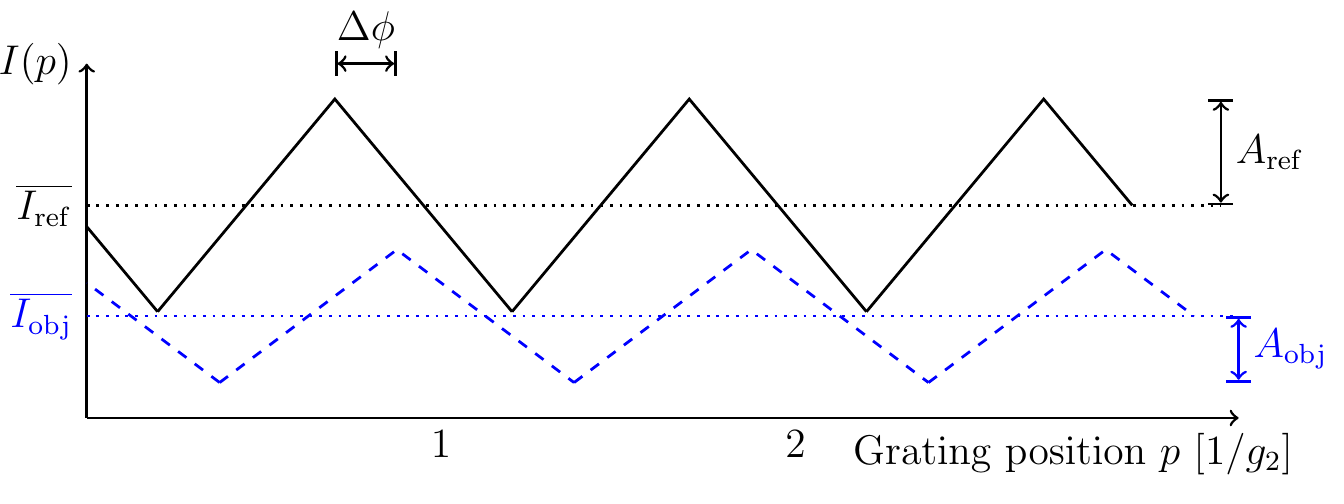}
	  \caption{Example intensity oscillations $I(p)$ detected in a detector pixel as a function of the lateral
	  position of grating G2. The solid line shows the reference oscillation without the object while the dashed line shows
	  the resulting oscillation with object. $\bar{I}$ are the mean values while $A$ are the amplitudes of the oscillations. $\Delta\phi$
	  is the phase difference between the oscillations and is the measurement parameter for the differential-phase
	  image.}
	  \label{fig:modulation}
	\end{center}
\end{figure}

Figure~\ref{fig:modulation} illustrates the intensity oscillations obtained in one pixel for an
ideal, monoenergetic case. The subscript ``ref'' refers to the reference oscillation without object, while ``obj''
denotes values with the object present in the beam. From these two curves the three image quantities of
grating-based X-ray phase-contrast imaging can be obtained: the absorption image as the ratio of the mean intensities
$\bar{I_\mathrm{obj}}\,/\,\bar{I_\mathrm{ref}}$ of the sampled oscillations, the differential-phase image
$\Delta\phi$ as the phase difference of the two intensity oscillations, and the dark field image as the ratio of the
visibilities $V_\mathrm{obj}\,/\,V_\mathrm{ref}$, with $V$ being defined as
\begin{equation}
	V = \frac{\imax-\imin}{\imax+\imin} \approx \frac{A}{\bar{I}}
\end{equation}
of the two oscillations, where \imax{} is the maximum and \imin{} the minimum detected in
a pixel. This can also be approximated as the fraction of the oscillation's amplitude $A$ and its mean intensity
$\bar{I}$. In the non-ideal case, where G0 has a not-infinitely-small slit width and a photon spectrum is used, the
triangular oscillations shown in Figure~\ref{fig:modulation} become sinusoidal due to the lack of spatial coherence generated by the source grating and the fact that the phase shift of G1 matches $\pi$ only for one energy -- the design energy
-- of the polychromatic X-ray spectrum.

\subsection{Phase retrieval}\label{sec:phase_retrieval}
From the two intensity oscillations $I(p)$ recorded as described in the previous section, the three image quantities can
be obtained in several ways. As the oscillations are sine functions, a fast Fourier transform~(FFT) can be performed for each pixel
for the reference- and the object oscillation.
Alternatively, a least-squares algorithm fitting a sine function to the data can be used. This technique is used for all
images shown in this work. The least-squares method has some advantages compared to the FFT method. For example, the
period of the sine function can be added as an additional fitting parameter to correct systematic sampling errors or
grating variations due to thermal expansion.
Furthermore, a change in the fit model function can correct for other systematic errors~\cite{weber11}.

Using one of the methods mentioned above, the parameters defining a sine function the offset $\bar I$, the
amplitude $A$ and the phase $\phi$ are determined from the measured data.
These are used to calculate the three image quantities in each pixel: the absorption $\mu = \frac{\bar{I_\mathrm{obj}}}{\bar{I_\mathrm{ref}}}$, the differential phase $\Delta \phi =
\phi_\mathrm{ref} - \phi_\mathrm{obj}$ and the dark field $D = \frac{V_\mathrm{obj}}{V_\mathrm{ref}}$.

\subsection{Setup parameters}\label{sec:setup}
All measurements presented were carried out at our benchtop phase-contrast imaging setup. It consists of a Siemens
MEGALIX X-ray tube driven at \SI{60}{kV} accelerator voltage. The anode material of the tube is tungsten and only the
intrinsic filtering was applied. For photon detection, the Varian PaxScan 2520D flat panel detector with {CsI} as
scintillation material and a pixel size of \SI{127\,\mathrm{x}\,127}{\mu m^2} was used. Before the measurements were
carried out, an offset and gain calibration of the detector was performed as recommended by the manufacturer.

\begin{table}
	\caption{\label{tab:gratings}Properties of the used gratings as stated on the delivery notes.}
	\begin{center}
	\begin{tabular}{@{}llll}
	\toprule
				&{source grating G0} & {phase grating G1} & {analyser grating G2}\\
				\midrule
	grating constant ($\mu\textrm{m}$) 		 & 23.95		 & 4.37		 &	2.40	\\ 
	thickness ($\mu\textrm{m}$) 		 &150			 & 8.7		 &100\\	
	duty cycle 						 &	 0.50	 &	 0.50	 &0.50\\   
	material 				     	 & {gold}		 & {nickel}	 &{gold} \\
	
%
	\bottomrule
	\end{tabular}
	\end{center}
\end{table}

The gratings of our setup were produced by Karlsruhe Institute of Technology~(KIT) employing the LIGA
method~\cite{reznikova08}. The parameters of the gratings are summarized in Table~\ref{tab:gratings}. For the
calculation of the fractional Talbot and grating distances a design energy of \SI{25}{keV} was
assumed. Except the source
grating G0, which had a quadratic field of view~(FOV) of \SI{5\,\mathrm{x}\,5}{cm^2}, all gratings were of 
rectangular shape with an effective area of \SI{2\,\mathrm{x}\,6}{cm^2}.

The measurements were carried out in the second constructive fractional Talbot distance ($n=2$ in
equation~\ref{eq:talbot1}).
Taking the source- and analyser-grating properties into account, the setup distances were calculated using
equation~\ref{eq:talbot1} and the theorem on intersecting lines resulting in \SI{161.2}{cm} for the G0-to-G1 distance
and \SI{15.9}{cm} for the G1-to-G2 distance.


The focal spot was located \SI{15.5}{cm} in front of the source grating G0, due to the size of the X-ray housing.
Further, the detector was placed \SI{5}{cm} behind the analyser grating, due to the dimensions of the grating mounting.
The object was placed \SI{10}{cm} in front of the phase grating G1.



\subsection{Animals}
Two C57BL/6 adult male mice were randomly selected from carbondioxide-killed ex-breeding stock at the
Franz-Penzold-Zentrum animal facility for investigation. Both animals were press fit in a \SI{50}{ml} conical
polyethylene tube to reduce any movement while the measurements were carried out. No further preparation procedures with
the samples were done.
This study complies to the institution's animal welfare and standard procedures.

\section{Results}
\begin{figure}
\begin{center}
  \includegraphics[width=.9\linewidth]{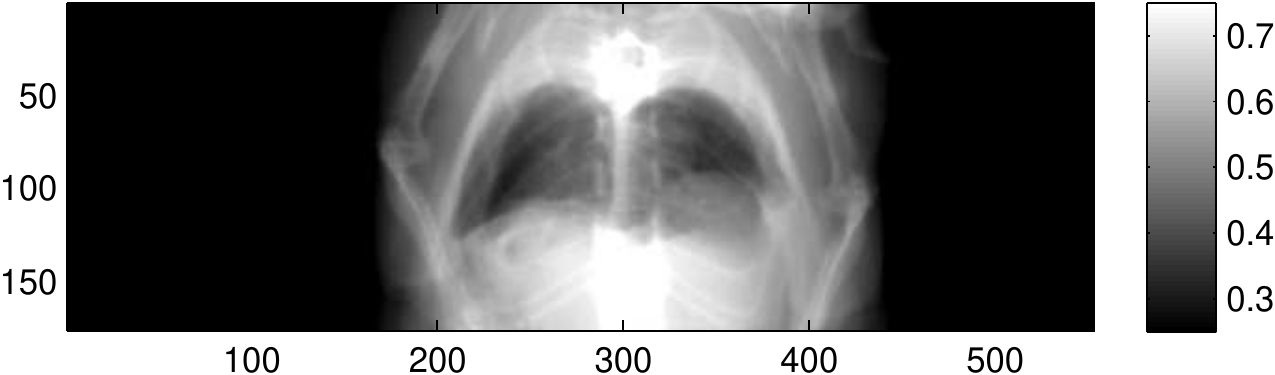}\\
  \includegraphics[width=.9\linewidth]{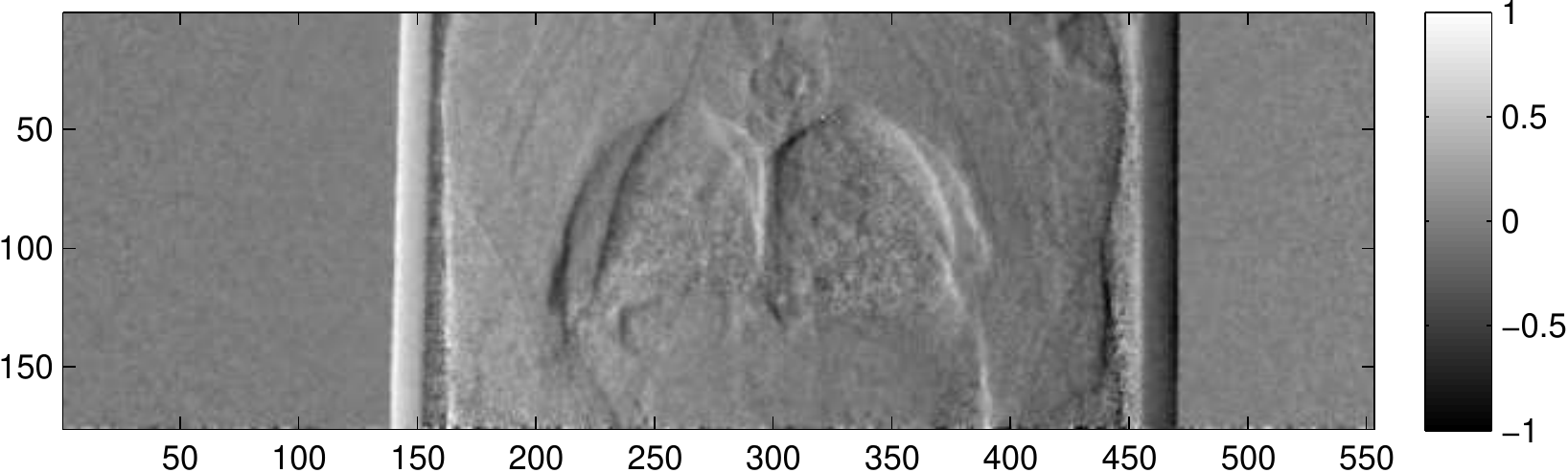}\\
  \includegraphics[width=.9\linewidth]{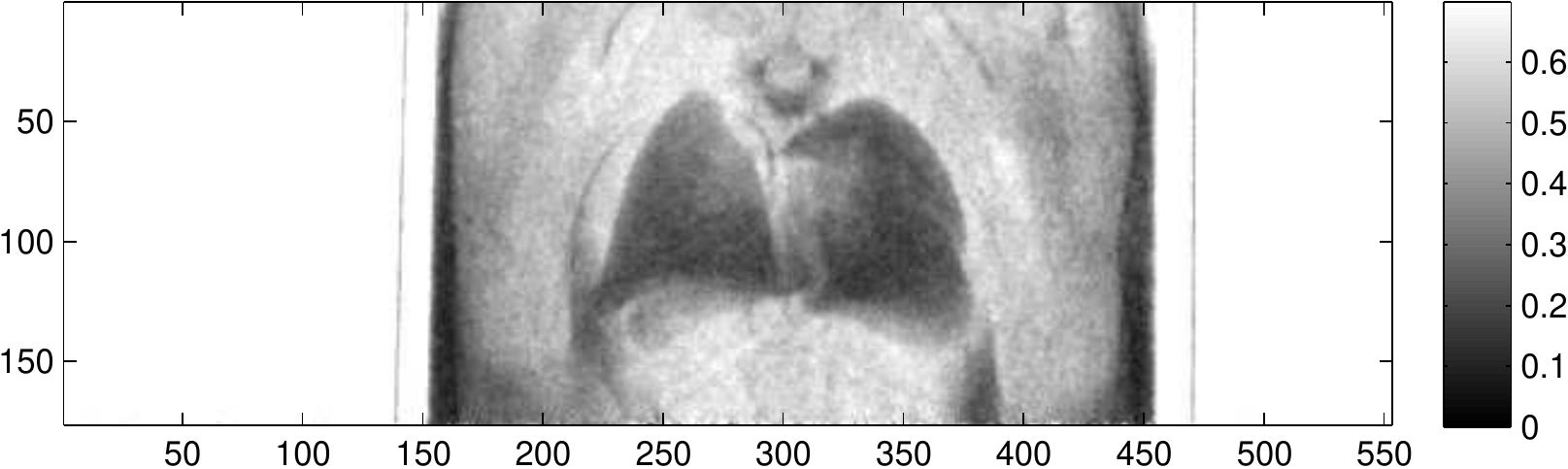}
  \caption{The three X-ray phase-contrast images of the lung region of a mouse in
  posterior-anterior (p.\,a.) orientation. Top: absorption image,  middle: differential phase image, bottom: dark
  field image.}
  \label{fig:lung_g}
\end{center}
\end{figure}
\emph{In situ} lung imaging of two healthy C57BL/6 adult mice was performed to obtain information on the signals of
phase-contrast imaging. Figure~\ref{fig:lung_g} shows the three images obtained of the lungs of a mouse in
posterior-anterior (p.a.) orientation, using eight phase steps with $\Delta x = \SI{0.3}{um}$ and an acquisition time of \SI{3.3}{s}
each at a tube current of \SI{30}{mA}. This corresponds to an air kerma of \SI{1.58}{mGy} measured at the object
position with a calibrated IBA Dosimax plus A HV dosimeter with the solid-state detector unit RQX 70kV.

\begin{figure}
\begin{center}
 {\includegraphics[width=.9\linewidth]{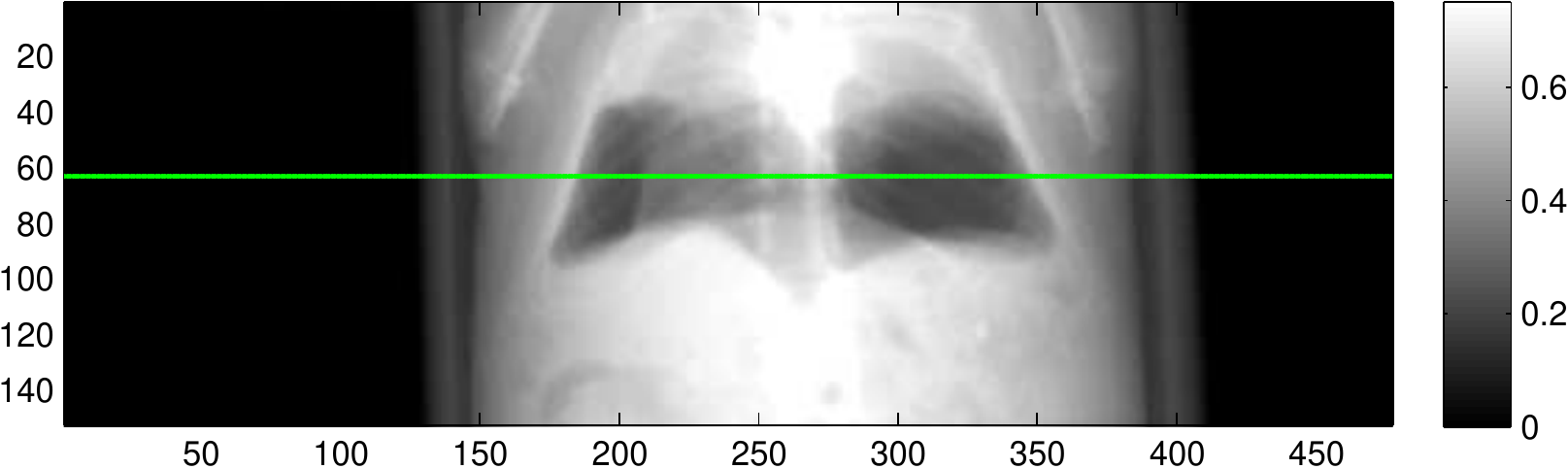}}\\
  {\includegraphics[width=.9\linewidth]{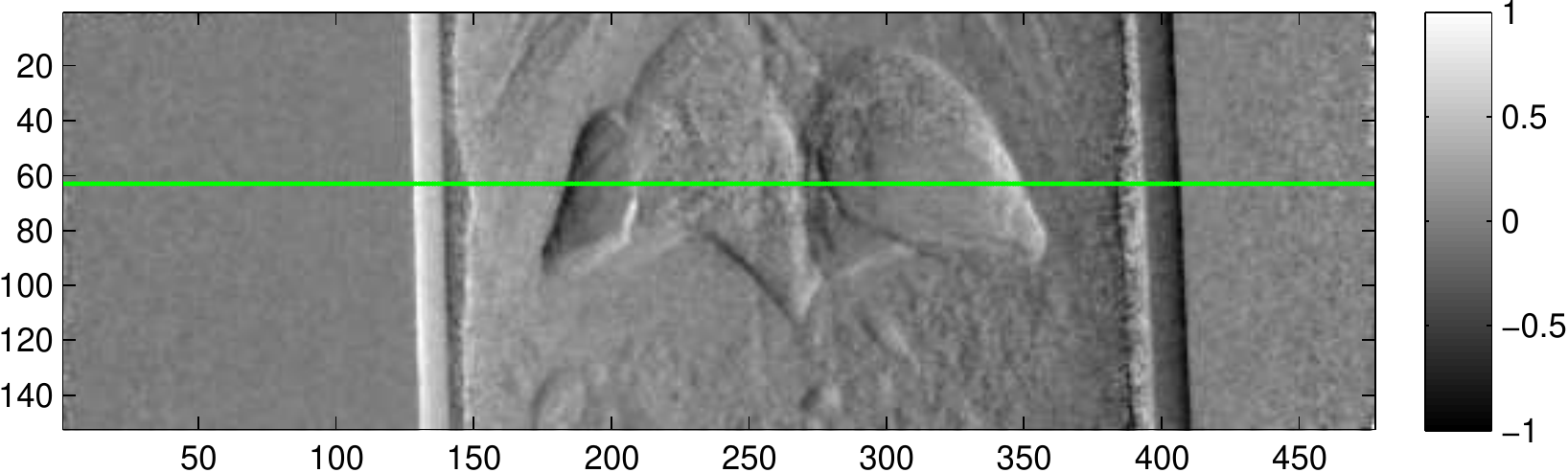}}\\
  {\includegraphics[width=.9\linewidth]{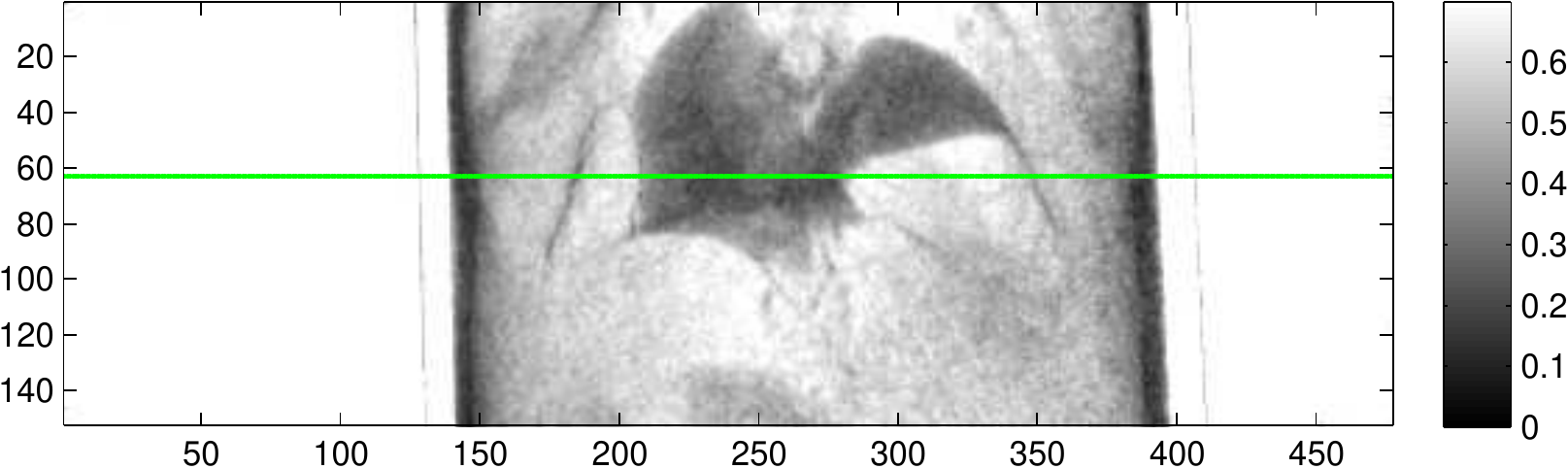}}
  \caption{Absorption (top), 
differential phase (middle) and dark field image (bottom) of the
lung region of a second mouse in posterior-anterior (p.\,a.\@) orientation. In the left part of the chest cavity (right
side of the images) a large space filled with air is visible, mainly in the dark field image. The horizontal line in
each image shows the location of the cross-section shown in
Figure~\protect\ref{fig:ct}.}
  \label{fig:lung_p}
\end{center}
\end{figure}

While the lung tissue is only weakly visible in the absorption image in Figure~\ref{fig:lung_g} the bony structure of
the mouse thorax can clearly be seen. In the differential-phase image the fur of the animal is clearly visible. Both
lungs as well as the spinal cord of the animal are shown.
The speckle noise in the phase image described by \citet{kitchen04} can be observed in the expected lung area.
Dark field imaging highlights clearly the fur-air interface as well as both lungs with
their typical anatomical structures and sizes.
Furthermore, the boundaries of the chest wall are clearly visible in the dark field and the differential-phase image.

These findings are supported by the images taken from the second mouse shown in Figure~\ref{fig:lung_p}. While in the
absorption image mainly the bones and the air-filled chest cavity can be seen, the other two
images  allow exact distinction between the empty chest cavity
and the lung tissue.



To evaluate the location of the lung tissue in the two mice computed tomographies~(CT) of them were performed.
This was done by rotating the object and acquiring 601 angular projections equally spaced over a full circle with the
setup described in Section~\ref{sec:setup}. A reference image was taken after every 15$^\mathrm{th}$ projection. This
leads to a total of \SI{4.7}{h} to acquire all projection images. For image reconstruction, a filtered
backprojection~(FBP) was applied using a ramp kernel for the absorption and the dark field data and a Hilbert kernel for
the phase data after performing the phase retrieval described in Section~\ref{sec:phase_retrieval}.

\begin{figure}
	\begin{center}
		{\includegraphics[width=.48\linewidth]{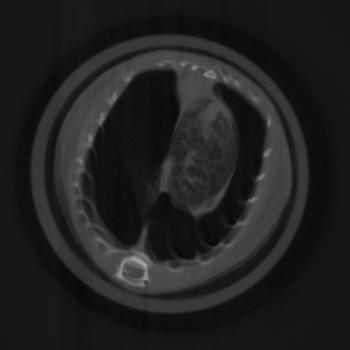}}\hfill
		{\includegraphics[width=.48\linewidth]{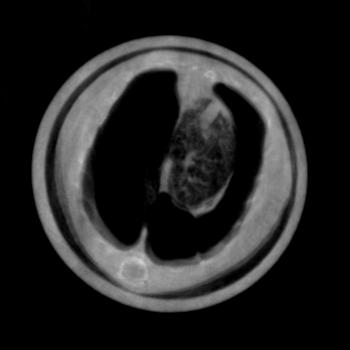}}\\
		{\includegraphics[width=.48\linewidth]{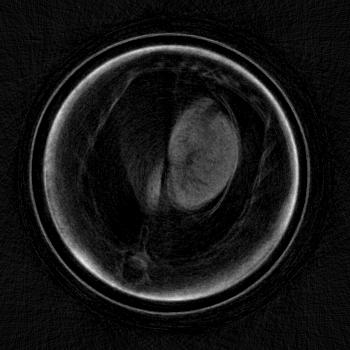}}
		\caption{Axial cross-sections of the second mouse obtained by
		phase-contrast computed tomography as marked in Figure~\ref{fig:lung_p}. Top left: absorption image,
		top right: phase image, and bottom: dark field image.	In this cross-section plane, lung tissue is
		only present in the right part of the mouse's chest cavity, while the rest is filled with air.}
		\label{fig:ct}
	\end{center}
\end{figure}

Exemplary for the data, one set of axial cross-sections of the lung region of the second mouse are displayed in
Figure~\ref{fig:ct} (marked as horizontal lines in Figure~\ref{fig:lung_p}).  In all three CT cross-sections, the lung
tissue can clearly be located in the chest of the mouse. The location found with the CT reconstruction corresponds with
the location visible in the dark field projection image and the speckled area in the
differential-phase image. In contrast, the exact position of the lung cannot be located
with the absorption image alone.

\section{Discussion}
The results obtained by the investigation of \emph{in situ} mouse lung tissue using grating-based X-ray phase-contrast imaging
are very encouraging.

First of all, three images containing different tissue properties are gained with one imaging process. This leads to an
increase in information from which radiologists can make a more profound diagnosis. An example for the complementarity
of information in the three images is shown in Figure~\ref{fig:lung_g}. There, the absorption image is dominated by the
bones of the animal and is therefore a good source for information about the skeletal structure and its diseases, like
bone fractures. In the differential-phase image on the other hand, the bones are suppressed.
Furthermore, the soft tissue contrast and the tissue edges are enhanced. This leads to a high organ visibility as in the
differential-phase image of Figure~\ref{fig:lung_g} the lungs and the upper airways can be seen. The third image
quantity, the dark field is a measure of the granularity of structures. Because of this, the lung, which consists of
many alveoli, produces a high dark field signal.

Comparing the three images, it becomes obvious that the differential phase and the dark field image complement the
absorption image and improve the identification of the shape of the lung. Especially in
cases where the granularity of the lung tissue is affected, the dark field image can lead to a faster diagnosis and
therefore treatment, as more information about the lung is accessible without a CT scan. It is imaginable for the
identification of pulmonary oedema, where the alveoli are filled with water. This leads to a more homogeneous view of the lung and can, in combination with the higher absorption of the water visible in the absorption image,
facilitate the diagnosis. Other examples for illnesses of the lung where the three different image quantities can
possibly be exploited are lung cancer, pneumonia or pneumothoraxes. For the last one, example images are shown in
Figure~\ref{fig:lung_p}. In the latter, the area which is occupied by the lung is just a very small part of the chest
cavity.
This area can clearly be identified in the dark field projection image, while in the absorption image the contrast
between space with and without lung tissue is hardly visible.

In addition to the new development of diagnostic tools for human health, this imaging approach opens a new avenue
to possibly reduce the number of animals required for some \emph{in vivo} studies. Long-term studies of
effects of toxic substances and both soft and bony tissue may be assessed \emph{in vivo} during the life of the animal.
Thus, a significantly smaller sample size in control and verum groups may be used and long-term studies, particularly in
academic institutions, become more feasible.

To quantify the location of lung tissue within the mouse, a CT scan of the second mouse was performed. Its results,
shown in Figure~\ref{fig:ct}, substantiate the above results that lung tissue is, indeed, only present in
the right part of the mouse's chest cavity in this cross-section plane.
It is imaginable that an assured diagnosis can be made without a computed tomography scan, which could lead
to a quickened diagnosis and a dose saving.


\section{Summary and Conclusion}

In conclusion, we presented our results on the investigation of \emph{in situ} lung tissue in mice obtained with
grating-based X-ray phase-contrast imaging.

The three different images show different aspects of the investigated sample. While the absorption
image shows mainly the skeletal structure, the differential-phase image enhances the tissue edges. This results in an
image with high soft-tissue contrast and suppressed bone signal. The dark field image is sensitive to granularity in the
sample.
Therefore, the lung can easily be distinguished from the surrounding tissue. Furthermore, we demonstrated that dark
field imaging can be used for the diagnosis of lung illnesses like a pneumothorax. For this case, we x-rayed
a mouse with a large volume of air in the chest cavity. In contrast to the conventional absorption image, the exact lung
location and size could easily be detected with only one projection image instead of a whole CT scan.


In the future, we will continue our investigations on mouse tissue, not only with respect to the lung, but also other
tissues, like the kidneys or cartilage. We will also apply this technology to investigate whether this approach can be
applied to long-term studies in a variety of disease-associated mouse studies to provide a new diagnostic tool and,
more importantly, to be able to reduce the number of animals needed for long-term interventions, slow healing models,
and organ toxicity and regenerative studies. We hope to find additional cases where grating-based X-ray phase-contrast
imaging yields an advantage compared to conventional X-ray imaging, not only to save dose, but also to increase the
speed of an assured diagnosis.

\begin{acknowledgments}
The authors want to thank the the German Federal Ministry of Education and Research for
funding this work within the PHACT project (BMBF 01 EZ 0923 DLR). Furthermore, we would like to thank Jasmin Horlitz of the
Franz-Penzoldt-Zentrum for the preparation of the mice.
\end{acknowledgments}


\begin{thebibliography}{16}%
\makeatletter
\providecommand \@ifxundefined [1]{%
 \@ifx{#1\undefined}
}%
\providecommand \@ifnum [1]{%
 \ifnum #1\expandafter \@firstoftwo
 \else \expandafter \@secondoftwo
 \fi
}%
\providecommand \@ifx [1]{%
 \ifx #1\expandafter \@firstoftwo
 \else \expandafter \@secondoftwo
 \fi
}%
\providecommand \natexlab [1]{#1}%
\providecommand \enquote  [1]{``#1''}%
\providecommand \bibnamefont  [1]{#1}%
\providecommand \bibfnamefont [1]{#1}%
\providecommand \citenamefont [1]{#1}%
\providecommand \href@noop [0]{\@secondoftwo}%
\providecommand \href [0]{\begingroup \@sanitize@url \@href}%
\providecommand \@href[1]{\@@startlink{#1}\@@href}%
\providecommand \@@href[1]{\endgroup#1\@@endlink}%
\providecommand \@sanitize@url [0]{\catcode `\\12\catcode `\$12\catcode
  `\&12\catcode `\#12\catcode `\^12\catcode `\_12\catcode `\%12\relax}%
\providecommand \@@startlink[1]{}%
\providecommand \@@endlink[0]{}%
\providecommand \url  [0]{\begingroup\@sanitize@url \@url }%
\providecommand \@url [1]{\endgroup\@href {#1}{\urlprefix }}%
\providecommand \urlprefix  [0]{URL }%
\providecommand \Eprint [0]{\href }%
\providecommand \doibase [0]{http://dx.doi.org/}%
\providecommand \selectlanguage [0]{\@gobble}%
\providecommand \bibinfo  [0]{\@secondoftwo}%
\providecommand \bibfield  [0]{\@secondoftwo}%
\providecommand \translation [1]{[#1]}%
\providecommand \BibitemOpen [0]{}%
\providecommand \bibitemStop [0]{}%
\providecommand \bibitemNoStop [0]{.\EOS\space}%
\providecommand \EOS [0]{\spacefactor3000\relax}%
\providecommand \BibitemShut  [1]{\csname bibitem#1\endcsname}%
\let\auto@bib@innerbib\@empty
\bibitem [{\citenamefont {Snigirev}\ \emph {et~al.}(1995)\citenamefont
  {Snigirev}, \citenamefont {Snigireva}, \citenamefont {Kohn}, \citenamefont
  {Kuznetsov},\ and\ \citenamefont {Schelokov}}]{snigirev95}%
  \BibitemOpen
  \bibfield  {author} {\bibinfo {author} {\bibfnamefont {A.}~\bibnamefont
  {Snigirev}}, \bibinfo {author} {\bibfnamefont {I.}~\bibnamefont {Snigireva}},
  \bibinfo {author} {\bibfnamefont {V.}~\bibnamefont {Kohn}}, \bibinfo {author}
  {\bibfnamefont {S.}~\bibnamefont {Kuznetsov}}, \ and\ \bibinfo {author}
  {\bibfnamefont {I.}~\bibnamefont {Schelokov}},\ }\bibfield  {title} {\enquote
  {\bibinfo {title} {On the possibilities of x-ray phase contrast microimaging
  by coherent high-energy synchrotron radiation},}\ }\href
  {http://dx.doi.org/10.1063/1.1146073} {\bibfield  {journal} {\bibinfo
  {journal} {Rev. Sci. Instrum.}\ }\textbf {\bibinfo {volume} {66}},\ \bibinfo
  {pages} {5486--5492} (\bibinfo {year} {1995})}\BibitemShut {NoStop}%
\bibitem [{\citenamefont {Wilkins}\ \emph {et~al.}(1996)\citenamefont
  {Wilkins}, \citenamefont {Gureyev}, \citenamefont {Gao}, \citenamefont
  {Pogany},\ and\ \citenamefont {Stevenson}}]{wilkins96}%
  \BibitemOpen
  \bibfield  {author} {\bibinfo {author} {\bibfnamefont {S.~W.}\ \bibnamefont
  {Wilkins}}, \bibinfo {author} {\bibfnamefont {T.~E.}\ \bibnamefont
  {Gureyev}}, \bibinfo {author} {\bibfnamefont {D.}~\bibnamefont {Gao}},
  \bibinfo {author} {\bibfnamefont {A.}~\bibnamefont {Pogany}}, \ and\ \bibinfo
  {author} {\bibfnamefont {A.~W.}\ \bibnamefont {Stevenson}},\ }\bibfield
  {title} {\enquote {\bibinfo {title} {Phase-contrast imaging using
  polychromatic hard x-rays},}\ }\href {http://dx.doi.org/10.1038/384335a0}
  {\bibfield  {journal} {\bibinfo  {journal} {Nature}\ }\textbf {\bibinfo
  {volume} {384}},\ \bibinfo {pages} {335--337} (\bibinfo {year}
  {1996})}\BibitemShut {NoStop}%
\bibitem [{\citenamefont {Chapman}(1997)}]{chapman97}%
  \BibitemOpen
  \bibfield  {author} {\bibinfo {author} {\bibfnamefont {D.}~\bibnamefont
  {Chapman}},\ }\bibfield  {title} {\enquote {\bibinfo {title} {Diffraction
  enhanced x-ray imaging},}\ }\href
  {http://dx.doi.org/10.1088/0031-9155/42/11/001} {\bibfield  {journal}
  {\bibinfo  {journal} {Phys. Med. Biol.}\ }\textbf {\bibinfo {volume} {42}},\
  \bibinfo {pages} {2015--2025} (\bibinfo {year} {1997})}\BibitemShut {NoStop}%
\bibitem [{\citenamefont {David}\ \emph {et~al.}(2002)\citenamefont {David},
  \citenamefont {N\"{o}hammer}, \citenamefont {Solak},\ and\ \citenamefont
  {Ziegler}}]{david02}%
  \BibitemOpen
  \bibfield  {author} {\bibinfo {author} {\bibfnamefont {C.}~\bibnamefont
  {David}}, \bibinfo {author} {\bibfnamefont {B.}~\bibnamefont {N\"{o}hammer}},
  \bibinfo {author} {\bibfnamefont {H.~H.}\ \bibnamefont {Solak}}, \ and\
  \bibinfo {author} {\bibfnamefont {E.}~\bibnamefont {Ziegler}},\ }\bibfield
  {title} {\enquote {\bibinfo {title} {Differential x-ray phase contrast
  imaging using a shearing interferometer},}\ }\href {\doibase
  10.1063/1.1516611} {\bibfield  {journal} {\bibinfo  {journal} {Applied
  Physics Letters}\ }\textbf {\bibinfo {volume} {81}},\ \bibinfo {pages}
  {3287--3289} (\bibinfo {year} {2002})}\BibitemShut {NoStop}%
\bibitem [{\citenamefont {Momose}\ \emph {et~al.}(2003)\citenamefont {Momose},
  \citenamefont {Kawamoto}, \citenamefont {Koyama}, \citenamefont {Hamaishi},
  \citenamefont {Takai},\ and\ \citenamefont {Suzuki}}]{momose03}%
  \BibitemOpen
  \bibfield  {author} {\bibinfo {author} {\bibfnamefont {Atsushi}\ \bibnamefont
  {Momose}}, \bibinfo {author} {\bibfnamefont {Shinya}\ \bibnamefont
  {Kawamoto}}, \bibinfo {author} {\bibfnamefont {Ichiro}\ \bibnamefont
  {Koyama}}, \bibinfo {author} {\bibfnamefont {Yoshitaka}\ \bibnamefont
  {Hamaishi}}, \bibinfo {author} {\bibfnamefont {Kengo}\ \bibnamefont {Takai}},
  \ and\ \bibinfo {author} {\bibfnamefont {Yoshio}\ \bibnamefont {Suzuki}},\
  }\bibfield  {title} {\enquote {\bibinfo {title} {Demonstration of x-ray
  talbot interferometry},}\ }\href {\doibase 10.1143/JJAP.42.L866} {\bibfield
  {journal} {\bibinfo  {journal} {Japanese Journal of Applied Physics}\
  }\textbf {\bibinfo {volume} {42}},\ \bibinfo {pages} {L866--L868} (\bibinfo
  {year} {2003})}\BibitemShut {NoStop}%
\bibitem [{\citenamefont {Weitkamp}\ \emph {et~al.}(2005)\citenamefont
  {Weitkamp}, \citenamefont {Diaz}, \citenamefont {David}, \citenamefont
  {Pfeiffer}, \citenamefont {Stampanoni}, \citenamefont {Cloetens},\ and\
  \citenamefont {Ziegler}}]{weitkamp05}%
  \BibitemOpen
  \bibfield  {author} {\bibinfo {author} {\bibfnamefont {Timm}\ \bibnamefont
  {Weitkamp}}, \bibinfo {author} {\bibfnamefont {Ana}\ \bibnamefont {Diaz}},
  \bibinfo {author} {\bibfnamefont {Christian}\ \bibnamefont {David}}, \bibinfo
  {author} {\bibfnamefont {Franz}\ \bibnamefont {Pfeiffer}}, \bibinfo {author}
  {\bibfnamefont {Marco}\ \bibnamefont {Stampanoni}}, \bibinfo {author}
  {\bibfnamefont {Peter}\ \bibnamefont {Cloetens}}, \ and\ \bibinfo {author}
  {\bibfnamefont {Eric}\ \bibnamefont {Ziegler}},\ }\bibfield  {title}
  {\enquote {\bibinfo {title} {X-ray phase imaging with a grating
  interferometer},}\ }\href {http://www.ncbi.nlm.nih.gov/pubmed/19498642}
  {\bibfield  {journal} {\bibinfo  {journal} {Optics express}\ }\textbf
  {\bibinfo {volume} {13}},\ \bibinfo {pages} {6296--304} (\bibinfo {year}
  {2005})}\BibitemShut {NoStop}%
\bibitem [{\citenamefont {Momose}\ \emph {et~al.}(2006)\citenamefont {Momose},
  \citenamefont {Yashiro}, \citenamefont {Takeda}, \citenamefont {Suzuki},\
  and\ \citenamefont {Hattori}}]{momose06}%
  \BibitemOpen
  \bibfield  {author} {\bibinfo {author} {\bibfnamefont {A.}~\bibnamefont
  {Momose}}, \bibinfo {author} {\bibfnamefont {W.}~\bibnamefont {Yashiro}},
  \bibinfo {author} {\bibfnamefont {Y.}~\bibnamefont {Takeda}}, \bibinfo
  {author} {\bibfnamefont {Y.}~\bibnamefont {Suzuki}}, \ and\ \bibinfo {author}
  {\bibfnamefont {T.}~\bibnamefont {Hattori}},\ }\bibfield  {title} {\enquote
  {\bibinfo {title} {Phase tomography by x-ray talbot interferometry for
  biological imaging},}\ }\href {http://dx.doi.org/10.1143/JJAP.45.5254}
  {\bibfield  {journal} {\bibinfo  {journal} {Japan J. Appl. Phys.}\ }\textbf
  {\bibinfo {volume} {45}},\ \bibinfo {pages} {5254--5262} (\bibinfo {year}
  {2006})}\BibitemShut {NoStop}%
\bibitem [{\citenamefont {Pfeiffer}\ \emph {et~al.}(2006)\citenamefont
  {Pfeiffer}, \citenamefont {Weitkamp}, \citenamefont {Bunk},\ and\
  \citenamefont {David}}]{pfeiffer06}%
  \BibitemOpen
  \bibfield  {author} {\bibinfo {author} {\bibfnamefont {F.}~\bibnamefont
  {Pfeiffer}}, \bibinfo {author} {\bibfnamefont {T.}~\bibnamefont {Weitkamp}},
  \bibinfo {author} {\bibfnamefont {O.}~\bibnamefont {Bunk}}, \ and\ \bibinfo
  {author} {\bibfnamefont {C.}~\bibnamefont {David}},\ }\bibfield  {title}
  {\enquote {\bibinfo {title} {Phase retrieval and differential phase-contrast
  imaging with low-brilliance x-ray sources},}\ }\href
  {http://dx.doi.org/10.1038/nphys265} {\bibfield  {journal} {\bibinfo
  {journal} {Nature Phys.}\ }\textbf {\bibinfo {volume} {2}},\ \bibinfo {pages}
  {258--261} (\bibinfo {year} {2006})}\BibitemShut {NoStop}%
\bibitem [{\citenamefont {Kitchen}\ \emph {et~al.}(2004)\citenamefont
  {Kitchen}, \citenamefont {Paganin}, \citenamefont {Lewis}, \citenamefont
  {Yagi}, \citenamefont {Uesugi},\ and\ \citenamefont {Mudie}}]{kitchen04}%
  \BibitemOpen
  \bibfield  {author} {\bibinfo {author} {\bibfnamefont {M~J}\ \bibnamefont
  {Kitchen}}, \bibinfo {author} {\bibfnamefont {D}~\bibnamefont {Paganin}},
  \bibinfo {author} {\bibfnamefont {R~A}\ \bibnamefont {Lewis}}, \bibinfo
  {author} {\bibfnamefont {N}~\bibnamefont {Yagi}}, \bibinfo {author}
  {\bibfnamefont {K}~\bibnamefont {Uesugi}}, \ and\ \bibinfo {author}
  {\bibfnamefont {S~T}\ \bibnamefont {Mudie}},\ }\bibfield  {title} {\enquote
  {\bibinfo {title} {{On the origin of speckle in x-ray phase contrast images
  of lung tissue}},}\ }\href {\doibase 10.1088/0031-9155/49/18/010} {\bibfield
  {journal} {\bibinfo  {journal} {Physics in Medicine and Biology}\ }\textbf
  {\bibinfo {volume} {49}},\ \bibinfo {pages} {4335--4348} (\bibinfo {year}
  {2004})}\BibitemShut {NoStop}%
\bibitem [{\citenamefont {Kitchen}\ \emph {et~al.}(2005)\citenamefont
  {Kitchen}, \citenamefont {Lewis}, \citenamefont {Yagi}, \citenamefont
  {Uesugi}, \citenamefont {Paganin}, \citenamefont {Hooper}, \citenamefont
  {Adams}, \citenamefont {Jureczek}, \citenamefont {Singh}, \citenamefont
  {Christensen}, \citenamefont {Hufton}, \citenamefont {Hall}, \citenamefont
  {Cheung},\ and\ \citenamefont {Pavlov}}]{kitchen05}%
  \BibitemOpen
  \bibfield  {author} {\bibinfo {author} {\bibfnamefont {M~J}\ \bibnamefont
  {Kitchen}}, \bibinfo {author} {\bibfnamefont {R~A}\ \bibnamefont {Lewis}},
  \bibinfo {author} {\bibfnamefont {N}~\bibnamefont {Yagi}}, \bibinfo {author}
  {\bibfnamefont {K}~\bibnamefont {Uesugi}}, \bibinfo {author} {\bibfnamefont
  {D}~\bibnamefont {Paganin}}, \bibinfo {author} {\bibfnamefont {S~B}\
  \bibnamefont {Hooper}}, \bibinfo {author} {\bibfnamefont {G}~\bibnamefont
  {Adams}}, \bibinfo {author} {\bibfnamefont {S}~\bibnamefont {Jureczek}},
  \bibinfo {author} {\bibfnamefont {J}~\bibnamefont {Singh}}, \bibinfo {author}
  {\bibfnamefont {C~R}\ \bibnamefont {Christensen}}, \bibinfo {author}
  {\bibfnamefont {a~P}\ \bibnamefont {Hufton}}, \bibinfo {author}
  {\bibfnamefont {C~J}\ \bibnamefont {Hall}}, \bibinfo {author} {\bibfnamefont
  {K~C}\ \bibnamefont {Cheung}}, \ and\ \bibinfo {author} {\bibfnamefont {K~M}\
  \bibnamefont {Pavlov}},\ }\bibfield  {title} {\enquote {\bibinfo {title}
  {{Phase contrast X-ray imaging of mice and rabbit lungs: a comparative
  study.}}}\ }\href {\doibase 10.1259/bjr/13024611} {\bibfield  {journal}
  {\bibinfo  {journal} {The British journal of radiology}\ }\textbf {\bibinfo
  {volume} {78}},\ \bibinfo {pages} {1018--27} (\bibinfo {year}
  {2005})}\BibitemShut {NoStop}%
\bibitem [{\citenamefont {Talbot}(1836)}]{talbot36}%
  \BibitemOpen
  \bibfield  {author} {\bibinfo {author} {\bibfnamefont {William Henry~Fox}\
  \bibnamefont {Talbot}},\ }\bibfield  {title} {\enquote {\bibinfo {title}
  {Facts relating to optical science},}\ }\href@noop {} {\bibfield  {journal}
  {\bibinfo  {journal} {Philos. Mag.}\ }\textbf {\bibinfo {volume} {9}},\
  \bibinfo {pages} {401--407} (\bibinfo {year} {1836})}\BibitemShut {NoStop}%
\bibitem [{\citenamefont {Suleski}(1997)}]{suleski97}%
  \BibitemOpen
  \bibfield  {author} {\bibinfo {author} {\bibfnamefont {T.J.}\ \bibnamefont
  {Suleski}},\ }\bibfield  {title} {\enquote {\bibinfo {title} {Generation of
  lohmann images from binary-phase talbot array illuminators},}\ }\href@noop {}
  {\bibfield  {journal} {\bibinfo  {journal} {Applied optics}\ }\textbf
  {\bibinfo {volume} {36}},\ \bibinfo {pages} {4686--4691} (\bibinfo {year}
  {1997})}\BibitemShut {NoStop}%
\bibitem [{\citenamefont {Morgan}(1982)}]{morgan82}%
  \BibitemOpen
  \bibfield  {author} {\bibinfo {author} {\bibfnamefont {CJ}~\bibnamefont
  {Morgan}},\ }\bibfield  {title} {\enquote {\bibinfo {title} {{Least-squares
  estimation in phase-measurement interferometry}},}\ }\href@noop {} {\bibfield
   {journal} {\bibinfo  {journal} {Optics Letters}\ }\textbf {\bibinfo {volume}
  {7}},\ \bibinfo {pages} {368--370} (\bibinfo {year} {1982})}\BibitemShut
  {NoStop}%
\bibitem [{\citenamefont {Creath}(1988)}]{creath88}%
  \BibitemOpen
  \bibfield  {author} {\bibinfo {author} {\bibfnamefont {K.}~\bibnamefont
  {Creath}},\ }\bibfield  {title} {\enquote {\bibinfo {title}
  {Phase-measurement interferometry techniques},}\ }\href@noop {} {\bibfield
  {journal} {\bibinfo  {journal} {Progress in optics}\ }\textbf {\bibinfo
  {volume} {26}},\ \bibinfo {pages} {349--393} (\bibinfo {year}
  {1988})}\BibitemShut {NoStop}%
\bibitem [{\citenamefont {Weber}\ \emph {et~al.}(2011)\citenamefont {Weber},
  \citenamefont {Bartl}, \citenamefont {Bayer}, \citenamefont {Durst},
  \citenamefont {Haas}, \citenamefont {Michel}, \citenamefont {Ritter},\ and\
  \citenamefont {Anton}}]{weber11}%
  \BibitemOpen
  \bibfield  {author} {\bibinfo {author} {\bibfnamefont {Thomas}\ \bibnamefont
  {Weber}}, \bibinfo {author} {\bibfnamefont {Peter}\ \bibnamefont {Bartl}},
  \bibinfo {author} {\bibfnamefont {Florian}\ \bibnamefont {Bayer}}, \bibinfo
  {author} {\bibfnamefont {Jurgen}\ \bibnamefont {Durst}}, \bibinfo {author}
  {\bibfnamefont {Wilhelm}\ \bibnamefont {Haas}}, \bibinfo {author}
  {\bibfnamefont {Thilo}\ \bibnamefont {Michel}}, \bibinfo {author}
  {\bibfnamefont {Andre}\ \bibnamefont {Ritter}}, \ and\ \bibinfo {author}
  {\bibfnamefont {Gisela}\ \bibnamefont {Anton}},\ }\bibfield  {title}
  {\enquote {\bibinfo {title} {Noise in x-ray grating-based phase-contrast
  imaging},}\ }\href {\doibase 10.1118/1.3592935} {\bibfield  {journal}
  {\bibinfo  {journal} {Medical Physics}\ }\textbf {\bibinfo {volume} {38}},\
  \bibinfo {pages} {4133--4140} (\bibinfo {year} {2011})}\BibitemShut {NoStop}%
\bibitem [{\citenamefont {Reznikova}\ \emph {et~al.}(2008)\citenamefont
  {Reznikova}, \citenamefont {Mohr}, \citenamefont {Boerner}, \citenamefont
  {Nazmov},\ and\ \citenamefont {Jakobs}}]{reznikova08}%
  \BibitemOpen
  \bibfield  {author} {\bibinfo {author} {\bibfnamefont {E.}~\bibnamefont
  {Reznikova}}, \bibinfo {author} {\bibfnamefont {J.}~\bibnamefont {Mohr}},
  \bibinfo {author} {\bibfnamefont {M.}~\bibnamefont {Boerner}}, \bibinfo
  {author} {\bibfnamefont {V.}~\bibnamefont {Nazmov}}, \ and\ \bibinfo {author}
  {\bibfnamefont {P.J.}\ \bibnamefont {Jakobs}},\ }\bibfield  {title} {\enquote
  {\bibinfo {title} {{Soft X-ray lithography of high aspect ratio SU8 submicron
  structures}},}\ }\href@noop {} {\bibfield  {journal} {\bibinfo  {journal}
  {Microsystem Technologies}\ }\textbf {\bibinfo {volume} {14}},\ \bibinfo
  {pages} {1683--1688} (\bibinfo {year} {2008})}\BibitemShut {NoStop}%
\end{thebibliography}
%

\end{document}